\documentstyle[12pt]{article}
\def\baselinestretch{1.2}
\def\ksection{\arabic{section}}

\def\@normalsize{\@setsize\normalsize{15pt}\xiipt\@xiipt
\abovedisplayskip 14pt plus3pt minus3pt%
\belowdisplayskip \abovedisplayskip
\abovedisplayshortskip  \z@ plus3pt%
\belowdisplayshortskip  7pt plus3.5pt minus0pt}
\def\small{\setsize\small{13.6pt}\xipt\@xipt
\abovedisplayskip 16pt plus3pt minus3pt%
\belowdisplayskip \abovedisplayskip
\abovedisplayshortskip  \z@ plus3pt%
\belowdisplayshortskip  7pt plus3.5pt minus0pt
\def\@listi{\parsep 4.5pt plus 2pt minus 1pt
            \itemsep \parsep
            \topsep 9pt plus 3pt minus 3pt}}
\def\underline#1{\relax\ifmmode\@@underline#1\else
	$\@@underline{\hbox{#1}}$\relax\fi}
\catcode`@=12
\catcode`\@=11
\def\thesection{\Roman{section}.}

\def\FERMIPUB{}
\def\FERMILABPub#1{\def\FERMIPUB{#1}}
\def\ps@headings{\def\@oddfoot{}\def\@evenfoot{}
\def\@oddhead{\hbox{}\hfill
	\makebox[.5\textwidth]{\raggedright
            \ignorespaces --\thepage{}--
	\hfill {\rm FERMILAB--Pub--\FERMIPUB}}}
\def\@evenhead{\@oddhead}
\def\subsectionmark##1{\markboth{##1}{}}
}
\ps@headings
\catcode`\@=12
\relax

\newcounter{appendix}
\def\appendix{\par
 \addtocounter{appendix}{1}
 \def\thesection{Appendix \Alph{appendix}:}
 \def\ksection{\Alph{appendix}}}

\newskip\humongous \humongous=0pt plus 1000pt minus 1000pt

\newif\ifdtup

\def\oldreffmt#1{\rlap{[#1]} \hbox to 2\parindent{}}

\def\figfmt#1{\rlap{Figure {#1}} \hbox to 1in{}}

\def\etal{\hbox{\it et al.}}

\def\VEV#1{\left\langle #1\right\rangle}
\def\slash#1{#1\!\!\!/\!\,\,}
\def\beq{\begin{equation}}
\def\eeq{\end{equation}}
\def\bea{\begin{eqnarray}}
\def\eea{\end{eqnarray}}
\def\half{\frac{1}{2}}

\def\bq{\begin{quote}}
\def\eq{\end{quote}}

\def\half{\frac{1}{2}}
\def \lta {\mathrel{\vcenter
     {\hbox{$<$}\nointerlineskip\hbox{$\sim$}}}}
\def \gta {\mathrel{\vcenter
     {\hbox{$>$}\nointerlineskip\hbox{$\sim$}}}}

\def \etal {{\it et al.}\ }
\relax
\begin{document}
\par \vskip .05in
\FERMILABPub{94/395--T}
\begin{titlepage}
\begin{flushright}
FERMILAB--PUB--94/395--T\\
November, 1994\\
Submitted to {\em Phys. Lett. B}
\end{flushright}
\vfill
\begin{center}
{\large \bf Topcolor Assisted Technicolor}
 \end{center}
  \par \vskip .1in \noindent
\begin{center}
{\bf Christopher T. Hill}
  \par \vskip .02in \noindent
{Fermi National Accelerator Laboratory\\
P.O. Box 500, Batavia, Illinois, 60510
\footnote{ Electronic address: (internet)
hill@fnal.fnal.gov
}}
  \par \vskip .02in \noindent
\end{center}
\begin{center}{\large Abstract}\end{center}
\par \vskip .01in
\begin{quote}
A condensate, $\overline{t}t$, arising from
$O(TeV)$ scale ``topcolor,'' in addition
to technicolor (and ETC) may naturally
explain the gauge hierarchy, the large
top quark mass, and contains
a rich system of testable consequences.
A triplet of strongly coupled
pseudo--Nambu--Goldstone bosons, ``top-pions,''
near the top mass scale is a generic prediction
of the models.  A new class of technicolor
schemes and associated phenomenology
is suggested in this approach.
 \end{quote}
 \par \vskip .02in \noindent

\vfill
\end{titlepage}
\def\baselinestretch{1.6}
\tiny
\normalsize

\vskip .1in
\noindent
{\bf I. Introduction and Synopsis}
\vskip .1in

\noindent
The large top quark mass is suggestive of
new dynamics associated with electroweak
symmetry breaking (ESB).
Top quark condensation models try to identify
all of the ESB with the formation of a
dynamical top quark mass.  In the fermion--loop
approximation one can write a simple Pagels--Stokar
formula which connects the Nambu--Goldstone boson
(longitudinal $W$ and $Z$)
decay constant, $f_\pi$, to the dynamical
mass, $m_c$ \cite{BHL} (we fix the normalization of
$f_\pi$ in eq.(7) below):
\beq
f_\pi^2 = \frac{N_c}{16\pi^2} m_{c}^2
( \log\frac{\Lambda^2}{m_c^2} + k)
\eeq
Here $m_c$ is the dynamical mass,
$k$ a constant
and $\Lambda$ the cut-off scale at which
the dynamical mass is rapidly going
to zero.
If electroweak symmetries are broken
dynamically by
the top quark mass, then $f_\pi = v_{wk} =(2\sqrt{2}G_F)^{-1/2}
\approx 175$ GeV,
and taking the cut--off  $\Lambda\sim 1.5$ TeV,
and $k\approx 1$,  we would
predict too large a top mass,
$m_c\sim 900 $ GeV.
Ergo, top condensation models must either
allow $\Lambda/m_t >>1$ with drastic
fine-tuning,
or invoke new dynamical mechanisms to try to obtain
a natural scheme.\footnote{In theories,
such as SUSY schemes,
in which the scale of new
physics may be large, $\Lambda\sim 10^{15} $ GeV,
the top quark mass
surprisingly saturates the Pagels--Stokar formula.
 In this case
$m_t$ is precisely determined by the
infra-red quasi-fixed point
\cite{FP}, which subsumes all
corrections to eq.(1).}

In this letter we wish to sketch
another possibility, which
seems to carry some intriguing implications.
We consider the possibility that: {\bf (i)}
electroweak interactions
are indeed broken by technicolor (TC)
\cite{TC} with an
extended technicolor (ETC) (yet,
one could replace these elements of
our discussion with Higgs scalars,
either as an approximation
to the TC/ETC dynamics, or as a
fundamental structure as in SUSY);
{\bf (ii)} the
top quark mass is large because
it is a combination
of a {\em dynamical condensate
component}, $(1-\epsilon)m_t$,
generated by a new strong dynamics,
together with a small {\em fundamental component},
 $\epsilon m_t$ (i.e, $\epsilon<<1$, generated by the
extended technicolor (ETC) or Higgs);
{\bf (iii)}
the new strong dynamics is assumed to
be chiral--critically strong but
spontaneously broken by TC  at the
scale $\sim 1$ TeV, and it is
coupled preferentially to the third
generation.
The new strong dynamics therefore
occurs primarily in interactions
that involve $\overline{t}t\overline{t}t$,
$\overline{t}t\overline{b}b$, and
$\overline{b}b\overline{b}b$,
while the ETC
interactions of the form $\overline{t}t\overline{Q}Q$,
are relatively
feeble.

Our basic assumptions, (i)-(iii),
leave little freedom of choice in
the new dynamics: We require a new
class of technicolor models
incorporating ``topcolor'' (TopC)
\cite{TopC}.\footnote{Else, we could
try to use the $SU(2)$ degrees of freedom
of the third generation, a possibility
which we will not consider
presently. }
In TopC the dynamics at the $\sim 1$ TeV scale
involves the following structure
(or a generalization thereof):
\beq
SU(3)_1\times SU(3)_2
\times U(1)_{Y1}\times U(1)_{Y2}
\times SU(2)_L \rightarrow
SU(3)_{QCD}\times U(1)_{EM}
\eeq
where
$SU(3)_1\times U(1)_{Y1}$
($SU(3)_2\times U(1)_{Y2}$)
generally couples preferentially
to the third (first and
second) generations.  The
$U(1)_{Yi}$
are just strongly rescaled
versions of
electroweak  $U(1)_{Y}$.  Hence we
are advocating a kind of
gauge group ``replication'' which
is generation sensitive.
$SU(3)_1\times U(1)_{Y1}$ is assumed
strong enough to form
chiral condensates which will
naturally be tilted in the top
quark direction by the $U(1)_{Y1}$ couplings.
This strong interaction is non-confining,
since the theory
spontaneously breaks down to
ordinary QCD$\times U(1)_{EM}$
at the TEV
scale by the technicolor gauge
group $G_{TC}$.
$U(1)_{Y1}$ is stronger than
the usual $U(1)_Y$, and there
need occur no significant
fine--tuning to arrange a
$\VEV{\overline{t}t}$
condensate, but not a
$\VEV{\overline{b}b}$ condensate,
by the simultaneous effects of
$SU(3)_1$ and $U(1)_{Y1}$
in the gap equation.
The $b$--quark mass
is then an interesting issue,
involving a combination
of ETC effects and instantons
in  $SU(3)_1$. The $\theta$--term
in $SU(3)_1$ may be the origin
of CKM CP--violation in these schemes.
Above all, the new spectroscopy
of such a system
should begin to materialize
indirectly
in the third generation
(e.g., in $Z\rightarrow \overline{b}b$)
or perhaps at the Tevatron in top
and bottom quark production.
A triplet of strongly coupled
pseudo--Nambu--Goldstone bosons
(PNGB's), $\tilde{\pi}^a$,
we dub ``top-pions,'' near the
top mass scale
is a generic prediction
of the models. The top-pions will
have a decay constant
of $f_\pi  \approx 50$ GeV, and a
strong coupling
given by a Goldberger--Trieman
relation,
$g_{tb\pi} \approx m_t/f_\pi\approx 3$,
potentially
observable in
$\tilde{\pi}^+\rightarrow t + \overline{b}$
if $m_{\tilde{\pi}} > m_t + m_b$\footnote{
Or the top quark may disappear into
a dominant decay mode
$t\rightarrow b +
(\tilde{\pi}^+ \rightarrow c+\overline{b})$
if $m_t > m_{\tilde{\pi}} + m_b$
and therefore has barely
been detected at the Tevatron.}.

\vskip .1in
\noindent
{\bf II. Topcolor Dynamics }
\vskip .1in
\noindent
We are relaxing the requirement
that a top
condensate account for the
full ESB and we are generalizing
the structure
in the interest in naturalness.
ESB can be primarily driven
by a technicolor group $G_{TC}$,
and/or TC can also provide
condensates which generate the
breaking of topcolor
to QCD and $U(1)_Y$.
The coupling constants (gauge
fields) of
$SU(3)_1\times SU(3)_2$  are
respectively
$h_1$ and $h_2$ ($A^A_{1\mu}$
and $A^A_{2\mu}$)
while for $U(1)_{Y1}\times U(1)_{Y2}$
they
are respectively  ${q}_1$ and $q_2$,
$(B_{1\mu}, B_{2\mu})$.
The $U(1)_{Yi}$ couplings are
then $q_i\frac{Y}{2}$, where $Y$
is usual electroweak
hypercharge.
A $(3,\overline{3})\times
(q_1,\overline{q}_2)$
techni--condensate breaks $SU(3)_1\times SU(3)_2
\times U(1)_{Y1}\times U(1)_{Y2}
\rightarrow SU(3)_{QCD}\times  U(1)_Y$
at a scale
$\Lambda \gta 240$ GeV, or it
fully breaks $SU(3)_1\times SU(3)_2
\times U(1)_{Y1}\times U(1)_{Y2}\times SU(2)_L
\rightarrow SU(3)_{QCD}\times  U(1)_{EM}$
at the scale $\Lambda_{TC}= 240$ GeV.
This typically
leaves a {\em residual global symmetry},
$SU(3)'\times U(1)'$,
implying a degenerate, massive
color octet of ``colorons,'' $B_\mu^A$,
and a singlet heavy
$Z'_{\mu}$.  The gluon $A_\mu^A$
and coloron $B_\mu^A$ (the SM $U(1)_Y$
field
$B_\mu$ and the $U(1)'$ field
$Z'_\mu$), are then defined by
orthogonal
rotations with mixing angle
$\theta$ [$\theta'$]:
\bea
& &
h_1\sin\theta = g_3;\qquad
h_2\cos\theta = g_3;\qquad
\cot\theta = h_1/h_2;\qquad
\frac{1}{g_3^2} = \frac{1}{h_1^2} +
 \frac{1}{h_2^2} ;
\nonumber \\
& &
q_1\sin\theta' = g_1;\qquad
q_2\cos\theta' = g_1;\qquad
\cot\theta' = q_1/q_2;\qquad
\frac{1}{g_1^2} = \frac{1}{q_1^2} +
 \frac{1}{q_2^2} ;
\eea
and $g_3$ ($g_1$) is the QCD ($U(1)_Y$)
coupling constant at $\Lambda_{TC}$.
We ultimately demand $\cot\theta \gg 1$
and  $\cot\theta' \gg 1$
to select the top quark direction for
 condensation.
The masses of the degenerate octet of
colorons and $Z'$ are given
by $M_B\approx g_3\Lambda/\sin\theta\cos\theta$
$ M_{Z'} \approx g_1\Lambda/\sin\theta'\cos\theta'$.
The usual QCD gluonic ($U(1)_Y$ electroweak)
interactions
are obtained for any quarks that carry
either $SU(3)_1$
or $SU(3)_2$ triplet quantum numbers
(or appropriately scaled $U(1)_i$ couplings).
 Integrating out $B$ and $Z'$
we obtain an effective low energy
four--fermion interaction:
\bea
{\cal{L}}' & =  &
-\frac{4\pi\kappa}{M_B^2}
\left[
\bar{t}\gamma_\mu
\frac{\lambda^A}{2} t +
 \bar{b}\gamma_\mu
\frac{\lambda^A}{2} b
\right]^2
 -\frac{4\pi
\kappa_{Y1}}{M_{Z'}^2}
\left[
\frac{1}{3}\bar{\psi}_L
\gamma_\mu \psi_L +
\frac{4}{3}\bar{t}_R
\gamma_\mu t_R
-\frac{2}{3}\bar{b}_R
\gamma_\mu  b_R
\right]^2
\nonumber \\
\eea
where $\psi_{L,R} = \half(1\pm\gamma^5)\psi$,
$\kappa = g_3^2\cot^2\theta/4\pi $
and
$\kappa_{Y1} = g_1^2\cot^2\theta'/4\pi $,
with cut-offs of $M_B$ and $M_{Z'}$.

The symmetry breaking leading to the top mass
is triggered by
the coloron and $Z'$ exchange
interactions, and can be
estimated in the NJL approximation.
For sufficiently large
$\kappa$ the attractive
four--fermion TopC interaction
would alone trigger formation of a
condensate, $\VEV{\overline{t}t + \overline{b}b}$, which
is globally custodially $SU(2)$
symmetric.
However, the $U(1)_{Y1}$ force is
attractive in the ${\overline{t}t}$
channel and repulsive in the
${\overline{b}b}$ channel.  Thus,
one obtains the pair of gap
equations for $m_t$
and $m_b$ ($M_{Z'}\approx
M_B$ for simplicity here):
\bea
m_t & = & \frac{3}{2\pi}
(\kappa + \frac{8}{27}\kappa_{Y1})m_t
\left(1 - \frac{m_t^2}{M_B^2}
\ln(M_B^2/m_t^2)\right)
\nonumber \\
m_b & = & \frac{3}{2\pi}
(\kappa - \frac{4}{27}\kappa_{Y1})
m_b\left(1 - \frac{m_b^2}{M_B^2}
\ln(M_B^2/m_t^2)\right)
\eea
Demanding nonvanishing $m_t$
and vanishing $m_b$, we require
critical and subcritical combinations:
\beq
\kappa + \frac{8}{27}\kappa_{Y1}
> \kappa_{crit} ;
\qquad
\kappa_{crit} > \kappa -
\frac{4}{27}\kappa_{Y1};
\qquad
(\kappa_{crit} = \frac{2\pi}{3}
\;\;\makebox{in NJL}).
\eeq
We can readily satisfy eqs.(6)
without fine--tuning.
Note that in the color singlet
channels the $U(1)_{Y1}$ effects
are actually $1/N_c$.

If  $M_{Z'}<< M_B$
then we should treat the
$U(1)_{Y1}$ as
a radiative enhancement
(suppression) of the $\overline{t}t$
($\overline{b}b$) channel.
Moreover, an analysis of the
full effective
Lagrangian reveals that one
obtains a composite $2$
Higgs--doublet model. One
doublet, $H_1$, couples to
$t_R$ and develops the
VEV; the other, $H_2$,
couples to $b_R$ and remains
a massive (non tachyonic) boundstate.
In the limit of
switching off $\kappa_{Y1}$, we
find that $H_1$ and $H_2$ form
a (custodial) $SU(2)_c$
doublet and the effective
Lagrangian is $SU(2)_c$ invariant.
The techniquarks ($Q_i$),
which have condensed by
the confining TC
interactions, have  acquired
constituent masses of order
$500$ GeV
and can be neglected on the
scales $\mu \sim m_t$ as well.
Thus,
$\VEV{\overline{Q}t}$
condensates, which
would break technicolor,
do not form.
Of course, the NJL approximation
is crude,
but as long as the
associated phase transitions
of the full strongly
coupled theory are approximately
second order, then analogous
rough--tuning in the full theory
should be possible.

Arranging that the couplings are
simultaneously large at $\sim 1$ TeV is
a further issue having to do
with a GUT scale boundary condition.
It suggests that low energy
couplings are small
because of the familiar
imbedding relations of eq.(3),
and GUT scale couplings are
larger than usually assumed.
Further strong dynamics
probably occurs in the ``desert''
(e.g. imbedding involving
$SU(2)_L$, etc.). Of course,
without knowing the
ETC theory $\sim 10^5$ GeV,
we cannot imagine reliable
extrapolations to the GUT scale.
In a theory like this we
are clearly {\em a priori}
abandoning the few
``successful
predictions'' of perturbative
(SUSY) unification.

ETC interactions (or fundamental Higgs)
generate the light fermion masses, and
give small contributions
to the $t$ and $b$ quark
masses as well.
The ETC masses are potentially
subject to resonant
enhancements in the full theory,
and without significant
fine--tuning we expect that the
largest fermion
mass scale that  ETC need
provide is $O(m_{c})\sim 1.0$
GeV
to $O(m_{s})\sim 0.1$ GeV,
\cite{Ap}.
As described below the $b$
quark receives
instanton contributions in
the gauge group
$SU(3)_1$.
Since ETC is required
to  generate $O(1.0)$ to $O(0.1)$
GeV masses,
it may need to be a
walking ETC \cite{WETC}.

Since the
top condensation is a spectator
to the TC (or Higgs) driven
ESB, there must
occur a multiplet of top-pions.
A chiral Lagrangian can be written:
\beq
L = i\overline{\psi}\slash{\partial}\psi - m_t(
\overline{\psi}_L\Sigma P\psi_R + h.c.) -\epsilon m_t
\overline{\psi}P\psi, \qquad
P=\left(\begin{array}{cc} 1 & 0\\ 0 & 0
\end{array}\right)
\eeq
and $\psi=(t,b)$, and $\Sigma = \exp(i\tilde{\pi}^a\tau^a
/\sqrt{2}f_\pi)$. Eq.(7) is invariant under
$\psi_L\rightarrow e^{i\theta^a\tau^a/2}\psi_L$,
$\tilde{\pi}^a\rightarrow \tilde{\pi}^a +
\theta^a f_\pi/\sqrt{2}$.
Hence, the relevant currents are left-handed,
$j_\mu^a = \psi_L\gamma_\mu\frac{\tau^a}{2}\psi_L$,
and $<\tilde{\pi}^a|j_\mu^b|0>
= \frac{f_\pi}{\sqrt{2}}p_\mu
\delta^{ab}$.  The Pagels-Stokar relation, eq.(1),
then follows by demanding that the
$\tilde{\pi}^a$ kinetic
term is generated by integrating
out the fermions.  The
top--pion decay constant estimated
from eq.(1) using
$\Lambda = M_{B}$ and $m_t = 175$
GeV is $f_\pi \approx 50$ GeV.
The couplings of the top-pions
take the form:
\beq
 \frac{m_t}{\sqrt{2}f_\pi}
\left[ {i}\overline{t}
\gamma^5 t \tilde{\pi}^0
+\frac{i}{\sqrt{2} }
\overline{t} (1-\gamma^5) b \tilde{\pi}^+
+ \frac{i}{\sqrt{2} }
\overline{b} (1+\gamma^5)t  \tilde{\pi}^-
\right]
\eeq
and the coupling strength is
governed by the relation
$g_{bt\tilde{\pi}} \approx m_t/\sqrt{2}f_\pi$.

The small ETC mass component of
the top quark implies that the masses
of the top-pions will depend
upon $\epsilon$ and $\Lambda$.
Estimating the induced top-pion
mass from the fermion loop yields
\cite{Schizon}:
\beq
m_{\tilde{\pi}}^2 =
\frac{N \epsilon m_t^2 M_B^2 }{8\pi^2 f_\pi^2}
= \frac{\epsilon M_B^2 }{\log(M_B/m_t)}
\eeq
where the Pagels-Stokar formula
is used for $f_\pi^2$
(with $k=0$) in the
last expression. For
$\epsilon = (0.03,\; 0.1)$,
$M_B\approx (1.5,\; 1.0) $ TeV,
and $m_t=180$ GeV  this
predicts $m_{\tilde{\pi}}= (180,\; 240)$ GeV.
The bare value of $\epsilon$
generated at the  ETC scale
$\Lambda_{ETC}$, however,
is subject to very large
radiative enhancements by
topcolor and $U(1)_{Y1}$ by
factors of order
$(\Lambda_{ETC}/M_B)^p \sim 10^1$,
where the $p\sim O(1)$.
Thus, we expect that
even a  bare value of
$\epsilon_0 \sim 0.005$
can produce sizeable $m_{\tilde{\pi}} > m_t$.
Note that $\tilde{\pi}$ will generally
receive gauge contributions
to it's mass; these are
at most electroweak in strength,
and therefore of
order $\sim 10$ GeV.

The $b$ quark receives
mass contributions from ETC of $O(1)$
GeV, but also an induced mass
from instantons in $SU(3)_{1}$.
The instanton
effective Lagrangian may be
approximated by the
`t Hooft flavor determinant (
we place the cut-off at $M_B$):
\beq
L_{eff} = \frac{k}{M_B^2}
e^{i\theta_1} \det
(\overline{q}_L q_R) + h.c.
= \frac{k}{M_B^2}
e^{i\theta_1}[(\overline{b}_L b_R)
(\overline{t}_Lt_R)
- (\overline{t}_L b_R)
(\overline{b}_Lt_R)] + h.c.
\eeq
where $\theta_1$ is the
$SU(3)_1$
strong $CP$--violation
phase. $\theta_1$
cannot be eliminated
because
of the ETC contribution
to the $t$ and $b$ masses.
It
can lead to
induced scalar couplings
of the neutral top--pion,
as in ref.\cite{Schizon},
and
an induced CKM CP--phase,
however, we will presently
neglect the effects of
$\theta_1$.

We generally expect $k\sim 1$
to $10^{-1}$ as in QCD.
Bosonizing in fermion bubble
approximation
$\overline{q}^i_Lt_R \sim
\frac{N}{8\pi^2} m_t M_B^2
\Sigma^i_1$, where
$\Sigma^i_j =
\exp(i\tilde{\pi}^a\tau^a/\sqrt{2}f_\pi)^i_j$ yields:
\beq
L_{eff} \rightarrow
\frac{Nk m_t}{8\pi^2}
e^{i\theta}[(\overline{b}_L b_R)
\Sigma^1_{1} +
(\overline{t}_Lb_R)
\Sigma^2_{1} + h.c.]
\eeq
This implies an instanton
induced $b$-quark mass:
\beq
m^\star_b \approx
\frac{3 k m_t}{8\pi^2} \sim 6.6\; k\; GeV
\eeq
This is not an unreasonable
estimate of the observed $b$ quark mass
as we might have feared it
 would be too large.
Expanding $\Sigma^i_j$,
there also occur induced
top--pion couplings to $b_R$:
\beq
\frac{m^\star_b}{\sqrt{2}f_\pi}
( i\overline{b} \gamma^5 b \tilde{\pi}^0
+
\frac{i}{\sqrt{2}}\overline{t}
(1+\gamma^5) b \tilde{\pi}^+
+ \frac{i}{\sqrt{2}}\overline{b}
(1-\gamma^5) t \tilde{\pi}^- )
\eeq

\noindent
{\bf III. Some Observables }
\vskip .1in
\noindent

The $t$ and $b$ quarks appearing
in, e.g., eq.(8),
are current--basis quarks. The
combination of TopC masses and
ETC masses yields a general
fermion mass matrix.
Diagonalization leads to  the
CKM matrix.
For the up-type (down-type)
quarks we take the
field redefinition to be
given by unitary matrices $U_{L,R}$
and $D_{L,R}$,
where the CKM matrix is
$V = U_L^\dagger D_L$. The leading
flavor changing
interactions involve then mixing to
the 2nd generation:
\bea
& &
 \frac{m_t}{\sqrt{2}f_\pi}
\left[ {i}\tilde{\pi}^0 (\overline{t}_R c_L U_{L,tc}
+
\overline{c}_R t_L U^*_{R,tc} )
+i{\sqrt{2} \tilde{\pi}^+ }
(\overline{t}_R s_L D_{L\;bs}
+\overline{c}_R b_L U^*_{Rtc})
+ h.c.
\right]
\eea
Exchange of top-pions
(as well as topgluons, $Z'$, and
the deeply bound $H_2$) generates
flavor changing effects.  By and
large we find that these can be
tolerably small in the low lying
states, up to the $B$ mesons,
but may show up in
processes like
$Z\rightarrow \overline{b}b$.

\vskip .1in
\noindent
{\em (i) $b\rightarrow s +\gamma:\; $ }
The top--pion  interactions
lead in principle to contributions
to the process $b\rightarrow s + \gamma$.
We estimate the ratio to the
SM result (we expect QCD
corrections to largely cancel):
\beq
\frac{B_{\tilde{\pi}}(s\rightarrow \gamma b)}{B_{SM}
(s\rightarrow \gamma b)}
\approx \left( 1 +
{\omega  }\right)^2
\qquad
\omega \approx
\left(\frac{D_{Lbs}v_{wk}}{V_{bs}f_\pi}
\right)^2 \frac{A(m_t^2/m_{\tilde{\pi}}^2)
}{3A(m_t^2/M_W^2)}
\eeq
In lowest order we have
the standard model contribution
plus the top-pion contribution
$C_7 = -\half A(m_t^2/M_W^2) -
(c)^2 A(m_t^2/m_{\tilde{\pi}}^2)/6$
and $c = D_{Lbs}v_{wk}
/V_{bs}f_\pi$ ($c$ is essentially
$\cot\beta$ in model I), comparing
eq.(14) to Grinstein \etal
\cite{Hew} eqs.(2.3, 2.29b).
For us,
$A(m_t^2/m_{\tilde{\pi}}^2)/3A(m_t^2/M_W^2)
 \approx 0.15$.
The SM result with QCD almost
saturates the observed
branching ratio. However, the
QCD corrections are very large,
and one cannot assume the NNLO
QCD effects are not also significant.
Conservatively, we might require,
${\omega}\lta 0.1$, hence,
$D_{L\;bs}/V_{bs}
\lta 0.2 $ using
$v_{wk}/{f_\pi}\sim 3.5$.
Since $D_{L\;bs}$ is
not measured (only the CKM element is)
this constraint
is not strictly binding. Identifying,
however,
$D_{L\;bs}$ with the corresponding
element in the square root of the
CKM matrix would favor
$\frac{D_{L\;bs}}{V_{bs}}\sim \half$,
the constraint
becomes slightly binding.
 We note that
the situation is not completely
settled \cite{Hew}.
There are, of course,
other apparently smaller
effects due to $Z'$,
$b$--coupled top-pions
from instantons,
and the deeply bound Higgs,
$H_1$ and $H_2$.

\vskip .1in
\noindent
{\em (ii) $\Delta S = 2$ and
$\Delta C = 2$  Effects: $\;\; $}
There occur FCNC effects induced
 by the CKM mixing
in the mass basis to the current
basis third generation.
In the current basis, we have
the neutral top--pion
coupled to the $t$ and $b$
quarks as
$i (m_t\overline{t}\gamma^5 t
+   m_b^\star\overline{b}\gamma^5 b)
{\pi^0}/{\sqrt{2}f_\pi}$.
Exchange of these neutrals
will induce $\Delta C= 2$
and  $\Delta S= 2$
effective interactions when
we rotate the $t$ and  $b$
quarks to their mass eigenbases,
$t \rightarrow t + O(\lambda)^2 c
+ O(\lambda)^3 u $ and
$ b \rightarrow b + O(\lambda)^2 s
+ O(\lambda)^3 d  $.
Thus, we obtain effective $\Delta C= 2$
and  $\Delta S= 2$ interactions:
\beq
\frac{m_t^2 O(\lambda^{10})}{2 m_{\tilde{\pi}}^2 f^2_\pi}
\overline{c}\gamma^5 u\overline{c}\gamma^5 u
+
\frac{m_b^2 O(\lambda^{10})}{2 m_{\tilde{\pi}}^2
f^2_\pi}\overline{s}\gamma^5
d\overline{s}\gamma^5 d
+ ...
\eeq
With $\lambda\sim O(10^{-1})$,
$m_{\tilde{\pi}} \gta m_t$,
these are of an acceptable
strength, e.g., in
comparison to
$(m_c^2 \lambda^2/ 128\pi^2 v^4_{wk})
\overline{s}\gamma^\mu d\overline{s}
\gamma_\mu d$.
Charged top--pions give box diagrams
of a similar strength.

\vskip .1in
\noindent
{\em (iii) $t\rightarrow \tilde{\pi}{}^+ + b$: $\;\;$}
The mode  $t\rightarrow \tilde{\pi}{}^+ + b$, if
kinematically allowed,
is ruled out if the top is seen
to have the conventional
rate $t\rightarrow W^+ + b$,
because the $\tilde{\pi}$ coupling is
very strong.  Small $m_{\tilde{\pi}}$
is disfavored by $b\rightarrow s+\gamma$
in any case.
{}From our perspective the observation of
a strongly coupled
$\tilde{\pi}{}^+ \rightarrow t + \overline{b}$
is a natural consequence of new
strong dynamics
associated with the generation of
the top quark mass.
The $\tilde{\pi}^+$ is expected to be a
broad state and
may be difficult to detect; the
$\tilde{\pi}^0$ may
be narrow if $m_{\tilde{\pi}} < 2 m_t$
and would decay
through anomalies to $gg$ and
$\gamma\gamma$,
(and to $\overline{b}b$ through
eq.(13)) and imitates
 some effects of states in
two-scale technicolor
(in contrast to \cite{EL} we do not
expect color octet
PNGB's associated with the
$f_\pi\sim 50$ GeV scale).

\vskip .1in
\noindent
{\em (v) $R_b$, $\sigma_{\overline{b}b}$,
$\sigma_{\overline{t}t}$:
}
It is particularly intriguing that,
while ETC interactions generally
lead to a suppression \cite{Chiv},
TopC schemes can contain significant
enhancements of
$R_b =\Gamma(Z\rightarrow \overline{b}b)/
\Gamma(Z\rightarrow \overline hadrons)$
\cite{Zhang}.  In the models
we have described both the topgluons
and the $Z'$ will enhance
$R_b$.  This is a desirable feature,
because when the observed LEP central value
for $R_b$ is fit
topgluons alone give too
much enhancement to top production at
the Tevatron \cite{HP,Zhang}. On the other hand
$Z'$ can
enhance $R_b$ with smaller impact upon
$\sigma_{\overline{t}t}$.  In our
present schemes we might expect
$M_{Z'}, M_{B} \sim 500 - 1000$ GeV
to accomodate acceptable observable
effects in top
production and $R_b$.  The $Z'$ may
then be observable in
$\sigma_{\overline{b}b}$ at the
Tevatron.
These potentially important effects, as
well as $S$, $T$ and $U$,
will be discussed in greater
detail elsewhere.

\vskip .1in
\noindent
{\bf IV. An Example of a New Model}
\vskip .1in
\noindent

We note that
a number of new models
is suggested by this approach.
In model
building we have several options:
(I) TC breaks both the EW interactions
and the
TopC interactions; (II) TC
breaks EW, and something else
breaks TopC;
(III) TC breaks only TopC and
something else drives ESB
(e.g., a fourth generation
condensate driven by TopC).
We presently show an example
of a very skeletal model in category (I)
in Table I.

For simplicity we
choose $G_{TC}=SU(3)_{TC1}\times
SU(3)_{TC2}$ and we have
indicated the
$U(1)_i$ hypercharge assignments.
The usual leptons
(and other techni--fields)
that are required to cancel
anomalies are not shown.
The techniquark condensate
$\VEV{\overline{Q}Q}$,  breaks
$ SU(3)_1  \times
SU(3)_2
\times U(1)_{Y1} \times
U(1)_{Y2}\rightarrow SU(3)\times U(1)_Y$,
but does not break
$SU(2)_L\times U(1)_Y$.
$SU(2)_L\times U(1)_Y\rightarrow
U(1)_{EM}$ occurs through the
condensate of
techniquarks $T_{L,R}$ which
feel the weaker $SU(3)_{TC2}\times U(1)_2$
interactions, thus $\VEV{\overline{T}T}$
is approximately custodially
$SU(2)$ invariant.
The third generation develops
the tilted condensate through the
$SU(3)_1\times U(1)_1$ interaction
with rough tuning of the tilting.
We have also assigned the second
generation
$(c,s)$ to the stronger $U(1)_1$
thus permitting a resonant
enhancement of the ETC mass scale
for charm and strange, so we
assume that the $U(1)_1$ coupling
is subcritical by itself.
The pattern suggests a further
$SU(3)_3$ replication for the
first generation.

We believe these models offer
new insights into the dynamical
origin of fermion masses and
electroweak symmetry breaking,
and merit further study.
Further model studies and
phenomenological applications
will be presented elsewhere.

\vskip .5in
\noindent
{\bf Acknowledgements}
\vskip .1in
\noindent
I thank Bill Bardeen, Estia Eichten and Ken Lane
for many insightful discussions.
This work was performed at the
Fermi National Accelerator Laboratory,
which is operated by Universities
Research Association, Inc., under
contract DE-AC02-76CHO3000 with
the U.S. Department of Energy.

\newpage
\vskip .1in
{
\vskip 0.2in
\begin{center}
\begin{tabular}{|| l | c | c | c | c | c | c | c ||}
\hline
field & $ SU(3)_{TC1} $  & $ SU(3)_{TC2} $
& $ SU(3)_1$  & $SU(3)_2$ & $ SU(2)_L$
& $ U(1)_{Y1}$  & $ U(1)_{Y2}$ \\ \hline
$Q_L$ & 3 & 1 & 3 & 1 & 1 &  $1$  & 0
\\ \hline
$Q_R$ & 3 & 1 & 1 &  3 & 1  & 0 & $1$
\\ \hline
$T_L=(T,B)_L$  & 1  & 3 & 1 & 1 & 2 & 0 &
 $\frac{1}{3}$
\\ \hline
$T_R=(T,B)_R$  & 1 & 3 & 1 & 1 & 1 & 0 &
 $(\frac{4}{3},\; -\frac{2}{3})$
\\ \hline
$t_L=(t,b)_L$  & 1 & 1 & 3 & 1 & 2  & $\frac{1}{3}$ & 0
\\ \hline
$t_R=(t,b)_R$  & 1  & 1 & 3 & 1 & 1  &
$(\frac{4}{3},\; -\frac{2}{3})$ & 0
\\ \hline
$c_L=(c,s)_L$   & 1 & 1 & 1 & 3 & 2 &
$\frac{1}{3}$  & 0
\\ \hline
$c_R=(c,s)_R$  & 1 & 1 & 1 & 3 & 1 &
 $(\frac{4}{3},\; -\frac{2}{3})$ & 0
 \\ \hline
$u_L=(u,d)_L$  & 1 & 1 & 1 & 3 & 2  & 0
& $\frac{1}{3}$ \\ \hline
$u_R=(u,d)_R$  & 1 & 1 & 1 & 3 & 1 & 0
& $(\frac{4}{3},\; -\frac{2}{3})$ \\ \hline
\hline
\end{tabular}
\end{center}
\begin{quote}
\vspace{.2in}
Table I: Gauge charge assignments
of quarks in a schematic model
$ SU(3)_{TC1} \times SU(3)_{TC2}
\times SU(3)_{1} \times SU(3)_{2}
\times SU(2)_L \times U(1)_{Y1}
\times
U(1)_{Y2}$.  Additional fields
(such as leptons) required
for anomaly cancellation and
are not shown. $\VEV{\overline{Q}Q }$
breaks
$SU(3)_{1} \times SU(3)_{2}
\times  U(1)_{Y1} \times U(1)_{Y2}
\rightarrow SU(3)\times U(1)_Y$,
and $\VEV{\overline{T}T }$ breaks
$SU(2)_{L}  \times  U(1)_{Y}
\rightarrow U(1)_{EM}$.
$\VEV{\overline{t}t }$
forms via
$SU(3)_1\times U(1)_{Y1}$.
\end{quote}}
\vskip 0.2in

\end{document}